\documentclass[sigconf]{acmart}
\AtBeginDocument{%
  }

\copyrightyear{2026}
\acmYear{2026}
\setcopyright{cc}
\setcctype{by-nc-nd}
\acmConference[Expanded 2026]{Expanded 2026: Conference on Animation and Interactive Art}{September 09--11, 2026}{Linz, Austria}
\acmBooktitle{Expanded 2026: Conference on Animation and Interactive Art (Expanded 2026), September 09--11, 2026, Linz, Austria}
\acmDOI{10.1145/3840416.3840461}
\acmISBN{979-8-4007-2865-5/2026/09}

\begin{document}

\title{Vertigo Vertigo: Reconstructing a Cinematic Ideal through its Predictive AI Double}


\author{Adam Cole}
\email{a.cole@arts.ac.uk}
\orcid{0000-0001-9715-314X}
\affiliation{%
  \institution{University of the Arts London}
  \city{London}
  \country{UK}
}

\author{Mick Grierson}
\email{m.grierson@arts.ac.uk}
\orcid{0000-0002-6981-5414}
\affiliation{%
  \institution{University of the Arts London}
  \city{London}
  \country{UK}
}


\begin{abstract}
\textit{Vertigo Vertigo} is a scene-for-scene AI reconstruction of Hitchcock's \textit{Vertigo} (1958), generated from only 2.78\% of the original film's frames. Using this sparse set of keyframe anchors, we perform first-last frame interpolation via a large video diffusion model to predict the intervening sequences. \textit{Vertigo} is itself a film about the obsessive reconstruction of an artificial ideal; \textit{Vertigo Vertigo} extends this logic to the material of the film, treating the canonical text as a probe for the normative conventions of classical cinema encoded within generative systems. Evaluated through computational analysis and critical feedback from media theorists (Lev Manovich, Shane Denson, Kevin L. Ferguson), the artifact demonstrates remarkable structural fidelity: 73.1\% of frames are recognizable as plausible renditions of \textit{Vertigo} and only 3.6\% fail catastrophically. This fidelity suggests that cinematic norms are deeply compressed within the model's latent priors. Aesthetically, the reconstruction is rendered as an unstable overlay between the original film and its predictive shadow, fueling a persistent doubt in the viewer's perception of authenticity --- a 21st-century vertigo. The work argues that generative media is not a paradigm shift from cinema but an acceleration of its logic of desire and false authenticity, extending from classical Hollywood through to the predictive media environments now reshaping contemporary perception.
\end{abstract}


\begin{teaserfigure}
  \includegraphics[width=\textwidth]{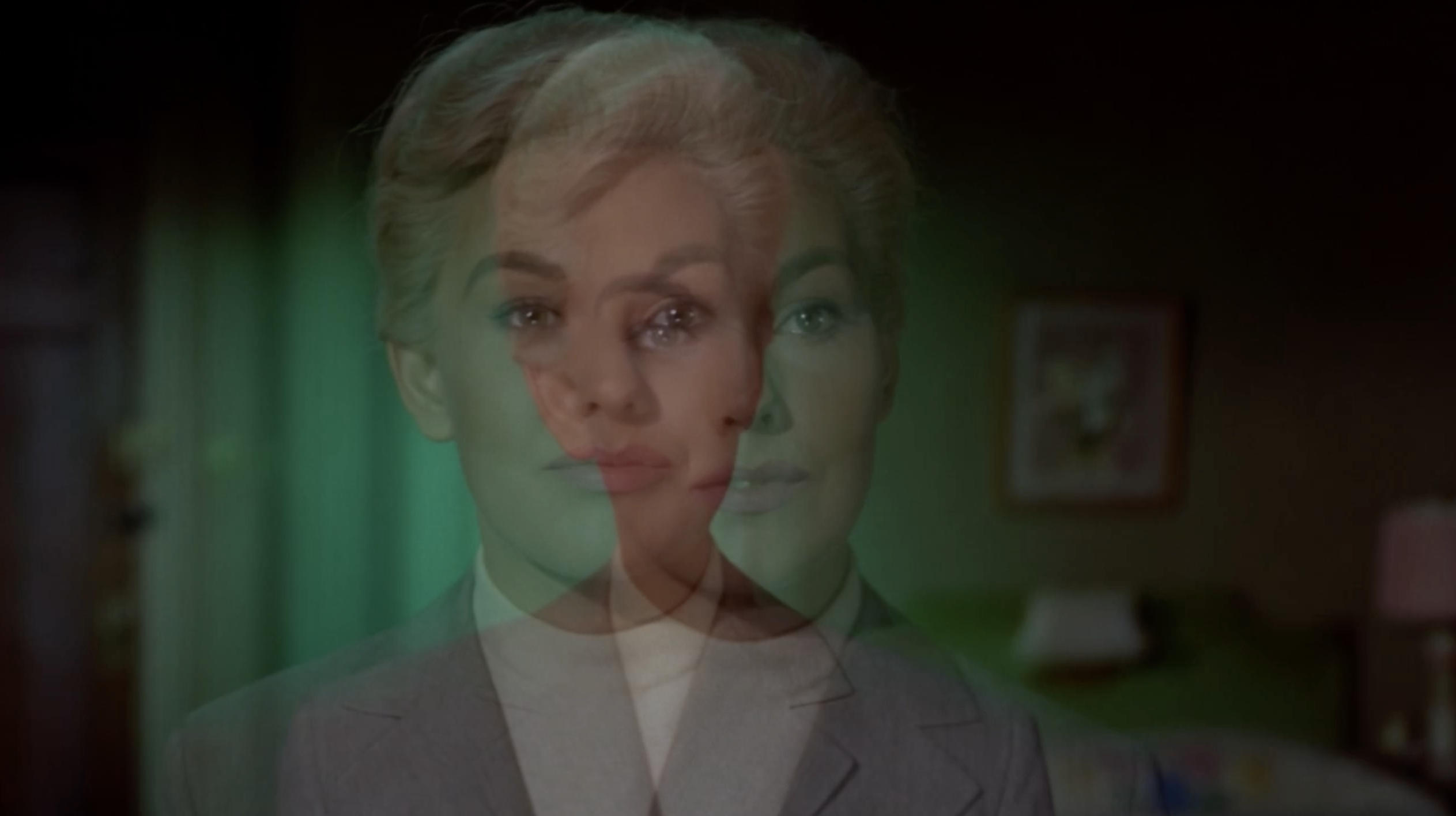}
  \caption{\textbf{\textit{Vertigo Vertigo} (2026)}: an AI reconstruction of Hitchcock's \textit{Vertigo} built from 2.78\% of the original film's frames, here showing the convergence-divergence relation between the original and its predictive shadow. Derived from: \textit{Vertigo} (1958), directed by Alfred Hitchcock. © Universal Studios.}
  \label{fig:teaser}
\end{teaserfigure}


\maketitle

\section{Introduction}
``I need you to be Madeleine for a while,'' Scottie famously demands of Judy in the final act of Alfred Hitchcock's \textit{Vertigo} (1958) \cite{hitchcockVertigo1958}. In this classic Hollywood film, a retired detective becomes dangerously obsessed with Madeleine, a woman he was hired to follow. After her apparent death, he remolds a new acquaintance, Judy, by meticulously contorting her appearance and behavior into the exact image of this lost love --- an image that, it turns out, was an artificial persona. This fixation on constructing a desired ideal finds a modern parallel in our interactions with generative AI: just as Scottie chases a perfected simulation, the contemporary culture of prompt-based AI, where one might literally type ``I need you to be'' into a machine, defines a relationship in which human desire and algorithmic conformity converge in the act of conception.

\textit{Vertigo Vertigo} is an experimental video artwork that plays with the original film's recursive logic of constructed identity and voyeuristic desire through a scene-for-scene AI remake of the original work. Utilizing a small fraction of the source footage, \textit{Vertigo Vertigo} applies this same logic of obsessive reconstruction to the material of the film itself. Specifically, the project algorithmically extracts keyframes amounting to 2.78\% of the source material and utilizes a 14-billion-parameter image-to-video diffusion model to run first-last frame interpolation between them, \textit{filling in the blanks} of the intermediary spaces. The resulting film, equal in length to \textit{Vertigo}, physically overlaps with the original at its keyframe anchor points but constantly diverges in the spaces between, creating an entirely hypothetical yet plausible rendition of the classic.

\textit{Vertigo} was chosen not merely for its thematic resonance with the project's method, but also because it sits at the intersection of three vital media art histories. First, \textit{Vertigo} features one of the first uses of computer-generated graphics in a feature film via John Whitney's title sequence animation \cite{youngbloodExpandedCinema1970}. Second, it is widely considered \textit{the cinematic ideal}, topping the 2012 \textit{Sight \& Sound} poll of the greatest films of all time and placing second in 2022 \cite{SightAndSound2022}. Finally, it serves as the foundational case study for normative gender representation in cinema, central to Laura Mulvey's articulation of the ``male gaze'' \cite{mulveyVisualPleasureNarrative1975}. The triangulation of these three qualities --- the computational image, the cinematic ideal, and the normativity of representation --- forms the foundation of this project.

\textit{Vertigo Vertigo} demonstrates the degree to which the visual and structural norms of classical cinema have been compressed into the latent priors of contemporary generative models, and exposes those compressed representations as an aesthetic friction the viewer can feel --- the discorrelation between the original and its predictive shadow. By processing a film whose narrative is itself about the obsessive reconstruction of an artificial ideal, the project reflects on a logic of desire and false authenticity that runs from classical cinema through to the present, where predictive, generative media environments increasingly reshape contemporary perception. The remainder of this paper situates the work within AI and film histories, details its methodology, presents a quantitative and qualitative analysis of the artifact, and draws on consultation with media theorists Lev Manovich, Shane Denson, and Kevin Ferguson to discuss the implications of this emerging paradigm.

\begin{figure*}[t]
    \centering
    \includegraphics[width=1\textwidth]{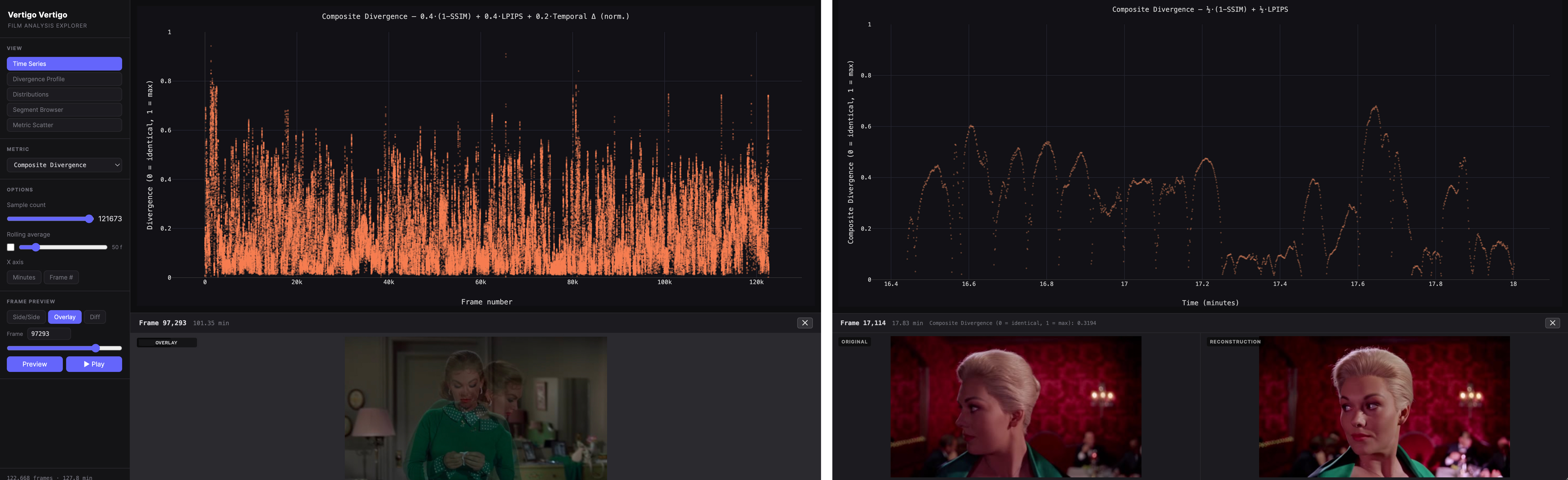}
    \caption{\textbf{Computational Analysis Suite.} \textbf{Left:} Composite divergence across all 122,668 frames of \textit{Vertigo Vertigo}. Distinct spikes correspond to atypical sequences (computer animation, dream sequences, low-light environments) where the model struggles to reconstruct the original. \textbf{Right:} Segment-level analysis of a single scene. Structural divergence drops to its lowest points at the extracted keyframe anchors and predictably peaks in the interpolated spaces between them. Embedded frames: \textit{Vertigo} (1958), directed by Alfred Hitchcock. © Universal Studios.}
    \label{fig:every-frame}
\end{figure*}

\section{Background}

\subsection{Generative Video and the Discorrelated Image}
Generative media continues to rapidly restructure the contemporary visual landscape, driven by advances in large video diffusion models such as Google Veo \cite{googleStateoftheartVideoImage2024} and open-weight systems like Wan \cite{wanteamWanOpenAdvanced2025}. These systems enable controllable pipelines via text-to-video, image-to-video, and video-to-video workflows, including the first-last frame interpolation we employ here, where a network is given two keyframe anchors and generates the sequence between them. As these moving-image affordances are integrated into the filmmaking pipeline \cite{zhangGenerativeAIFilm}, from professional studios to social media creators and experimental video artists, the line between camera-recorded \textit{reality} and computer-generated \textit{prediction} is eroding. This transformation carries immense industrial, political, and cultural stakes, as evidenced by the 2023 SAG-AFTRA strikes regarding the ownership and algorithmic replication of human likeness \cite{yamazakiDigitalReplicasDemocracy2024}.

The production and distribution of this AI-generated imagery fall neatly within Shane Denson's framework of \textit{discorrelated images} \cite{densonDiscorrelatedImages2020}. These are images that ``do not correlate with the abilities and limits of human perception,'' displacing the traditional spatial and temporal relationships viewers hold with conventional analog cinema. Discorrelation goes beyond technical classification; it is an affective, experiential change in how viewers relate to a media environment that now operates beyond human perceptual space-time. Contemporary AI video also materializes a shift Lev Manovich anticipated in \textit{The Language of New Media} \cite{manovichLanguageNewMedia2001}, where he described a paradigm in which live-action filmmaking is increasingly subsumed by the logic of computer animation. \textit{Vertigo Vertigo} materializes both frameworks, visualizing a discorrelated reality in which the past is reanimated entirely through the predictive models of the present.

\subsection{Reconstructing Hitchcock: A Tradition}
Because Hitchcock's films, and \textit{Vertigo} in particular, are both representative and constitutive of classical cinema's ideals and its normative representations, they have long served as the raw material for technological and structuralist adaptation. There is a rich lineage of technologically-mediated reinterpretations of this canon: Douglas Gordon's \textit{24 Hour Psycho} (1993) \cite{gordon24HourPsycho1993} used analog video to radically decelerate the image; Gus Van Sant's \textit{Psycho} (1998) \cite{santPsycho1998} utilized the shot-for-shot remake as a subtly queer structuralist experiment; Chris Marker's \textit{Sans Soleil} (1983) \cite{markerSansSoleil1983} visited the geographical locations of \textit{Vertigo} as an architecture for retracing the failures of human memory; and Grégory Chatonsky's \textit{Vertigo@home} (2007/2015) \cite{gregorychatonskyVertigohome2007} explored this territory in the computational era by retracing Scottie's routes via Google Street View.

Within the specific realm of AI film reconstruction, Terence Broad's \textit{Blade Runner---Autoencoded} (2016) \cite{broadAutoencodingBladeRunner2017} serves as a foundational precedent, demonstrating the knowledge-producing potential of passing a cultural text through a neural bottleneck. Kevin L. Ferguson's \textit{Volumetric Cinema} \cite{fergusonVolumetricCinema2015} and Daniel Chávez Heras's \textit{Cinema and Machine Vision} \cite{chavezherasCinemaMachineVision2024} bookend a critical-technical tradition that uses computational processing to reveal patterns within classical film and, inversely, what such processing reveals about algorithmic vision itself. More recently, Rachel Maclean's deepfake short \textit{DUCK} (2024) \cite{macleanDuck2024} reanimates Sean Connery and Marilyn Monroe to interrogate the gendered iconography of classical Hollywood, situating AI within a longer tradition of artists who appropriate cinematic archetypes to expose the construction of the feminine ideal.

While \textit{Vertigo Vertigo} synthesizes the structuralist instinct of these art historical precedents with the AI translation of earlier neural experiments, it remains distinct from both. This project utilizes a contemporary diffusion-based video model conditioned on predicting content from sparse keyframes. Unlike previous generative experiments that process films frame-by-frame through neural translation or scene-by-scene through generative prompts, this project operates on the structural skeleton of the entire feature film, producing a measurable, oscillating friction between an original ideal and algorithmic prediction that reflects our specific technological moment. To our knowledge, no prior work has measured continuous divergence across a full feature-length AI reconstruction at frame-level granularity, nor connected such measurement to the inheritance of cinematic conventions within generative systems.

\section{Methodology}

\subsection{Generation Pipeline}

Sparse-keyframe interpolation was chosen as the experimental method because it isolates the model's predictive potential without semantic injection. No scene-specific text prompts guide the process; instead, the model must predict motion and structure from minimal keyframe anchors. The 2.78\% dataset\footnote{The \textit{Vertigo} (1958) \cite{hitchcockVertigo1958} source material is used under UK fair dealing provisions for computational research and critical review.} is sparse enough to require the model to draw on its broader visual priors, but dense enough to prevent overly long sequences beyond the model's generation abilities. What the model produces in the spaces between is therefore a measure of how well the model can approximate a feature film, given only its skeleton.

To produce the artifact, we developed a custom AI video reconstruction system using Wan 2.2, a 14-billion-parameter image-to-video diffusion model. The pipeline operates in three stages. (1) \textbf{Segmentation}: the source film is segmented by extracting keyframes at every hard cut; for longer takes, additional keyframes are sampled every 5--6 seconds, with each segment constrained to around 81 frames, the model's optimal generation length. (2) \textbf{Generation}: the model performs first-last frame interpolation for each segment ($1280\times720$ resolution at $16$fps), generating the frames between the two anchor keyframes without segment-specific prompting. (3) \textbf{Combination:} the resulting clips are sequentially joined and trimmed to match the original film's length, ensuring the reconstruction aligns with the source at every keyframe and drifts into algorithmic prediction in the spaces between. 

The complete pipeline produced 2,126 segments anchored by 3,416 unique keyframes: the 2.78\% of the source material the model was given to work from.

\begin{figure*}[t]
    \centering
    \includegraphics[width=1\linewidth]{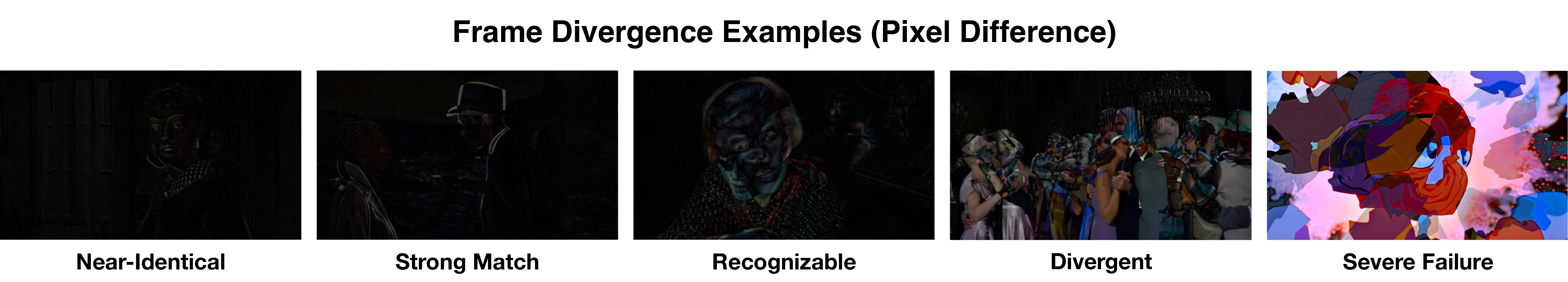}
    \caption{Representative frames for each tier of reconstruction quality using frame-wise pixel difference, where darker regions indicate close agreement between original and reconstruction and more colorful regions show greater divergence. Derived from: \textit{Vertigo} (1958), directed by Alfred Hitchcock. © Universal Studios.}
    \label{fig:tier-examples}
\end{figure*}

\subsection{Modes of Presentation}

The reconstruction is presented in three display modes. \textbf{Overlay}, the primary artwork, layers the reconstruction onto the original at 50\% opacity, producing a rhythmic double-exposure: convergence at the keyframe anchors, ghostly divergence between them. \textbf{Side-by-Side} displays the two films simultaneously for direct structural comparison. \textbf{Difference} renders the pixel-wise absolute difference between original and reconstruction, isolating the AI's deviations as abstract visual data.

\section{The Artifact and Evaluation}

The artifact was evaluated through a combination of computational analysis and qualitative consultation with media theorists specializing in post-cinema, computational aesthetics, and film studies.

For the qualitative consultation, each theorist was sent sample video clips representing the three display modes and provided access to the full reconstructed film. Feedback was structured around three open-ended questions on the legibility of the intervention, the relationship between the project's method and \textit{Vertigo}'s themes of voyeurism and the feminine ideal, and the implications of the machine's predictive disposition for inherited cinematic conventions. Responses were collected via email correspondence and inform the Discussion that follows.

\subsection{Metrics of Divergence}

To evaluate the reconstruction, we measured per-frame structural similarity (SSIM, capturing lower-level signals from luminance and contrast), perceptual similarity (LPIPS, using neural networks to capture higher-level perceptual features), and temporal change between consecutive frames, alongside a weighted composite of all three. A custom interactive analysis suite plots every frame from 0 (identical) to 1 (completely divergent), shown in Fig.~\ref{fig:every-frame}.

\subsection{Quantitative Analysis}
\subsubsection{Frame-Level: Measuring Divergence}
Because no single metric captures the full character of divergence, the frames are partitioned into tiers defined by joint thresholds across SSIM (structural) and LPIPS (perceptual). The cutoffs correspond to recognizable visual behaviors, from near-identical reconstruction to severe structural and semantic failure (Fig.~\ref{fig:tier-examples}).

\begin{figure}[b]
    \centering
    \includegraphics[width=1\linewidth]{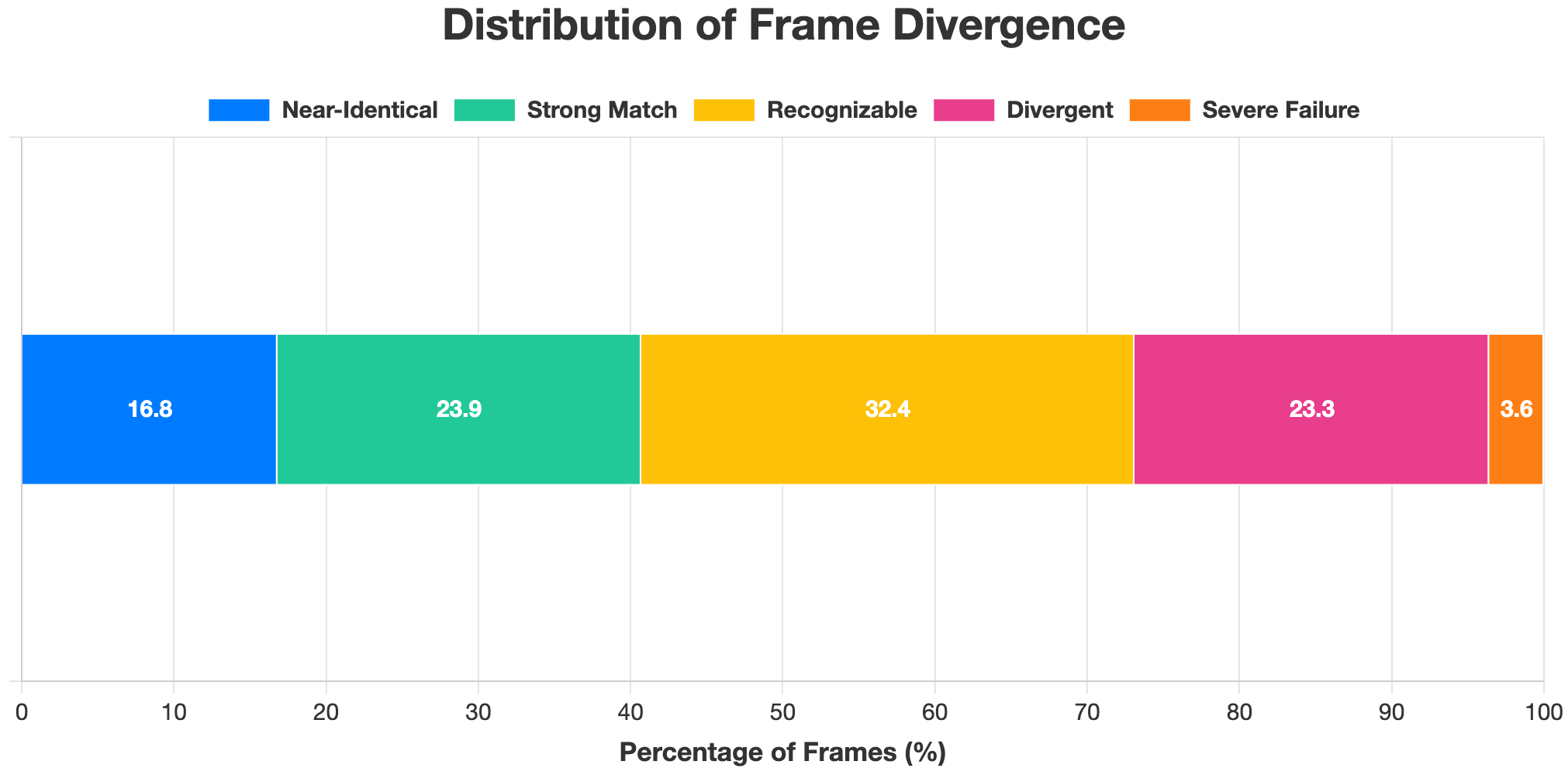}
    \caption{Frame-level reconstruction quality across all frames of the AI-reconstructed \textit{Vertigo}. The modal tier is Recognizable at 32.4\%, with the upper three tiers together accounting for 73.1\% of frames. Only 3.6\% of frames are severe failures.}
    \label{fig:tier-distribution}
\end{figure}

With 16.8\% of frames \textit{near-identical} and 3.6\% \textit{severe failures}, the bulk of the film resides in the middle: a plausible rendition of \textit{Vertigo}, yet perceptually distinct (Fig.~\ref{fig:tier-distribution}). That 73.1\% of frames remain recognizable demonstrates that classical cinema's visual and structural conventions are substantially retrievable from a small fraction of frames, suggesting they are deeply encoded within the priors of contemporary generative models.

\subsubsection{Segment-Level: Oscillating Rhythm} Evaluating the film at the segment level, we can see the signature of the generation method: near identical frames at the keyframes, and significant divergence between them (Fig.~\ref{fig:every-frame}). Comparing divergence across segments, we see a consistent U-pattern, confirming the reconstruction is measurably most faithful at keyframe anchors and most free between them. Notably, divergence depth is not driven by interpolation distance. The correlation between segment length and U-depth is \textit{r} = 0.036 (essentially zero). The reconstruction's content, not its temporal scale, governs its fidelity, raising the question of what kinds of images the model has been trained to produce most fluently.

\subsubsection{Scene-Level: Spikes of Divergence} Reviewing the distribution of the scatter plot (Fig.~\ref{fig:every-frame}), we see a few spikes that correspond to specific scenes with remarkably more divergence than the rest of the film. These are the John Whitney animated opening sequence (frames 0--3k), the surreal dream sequence (frames 80--83k), and a few sequences at the end in lower lighting (frames 120--122k). 

\subsection{Qualitative Evaluation}

While the quantitative data reveals the overall structural fidelity, a closer qualitative reading of the model's outputs explores the project's aesthetic potential.

\subsubsection{Where the Model Fails}

\begin{figure}[h!]
    \centering
    \includegraphics[width=1\linewidth]{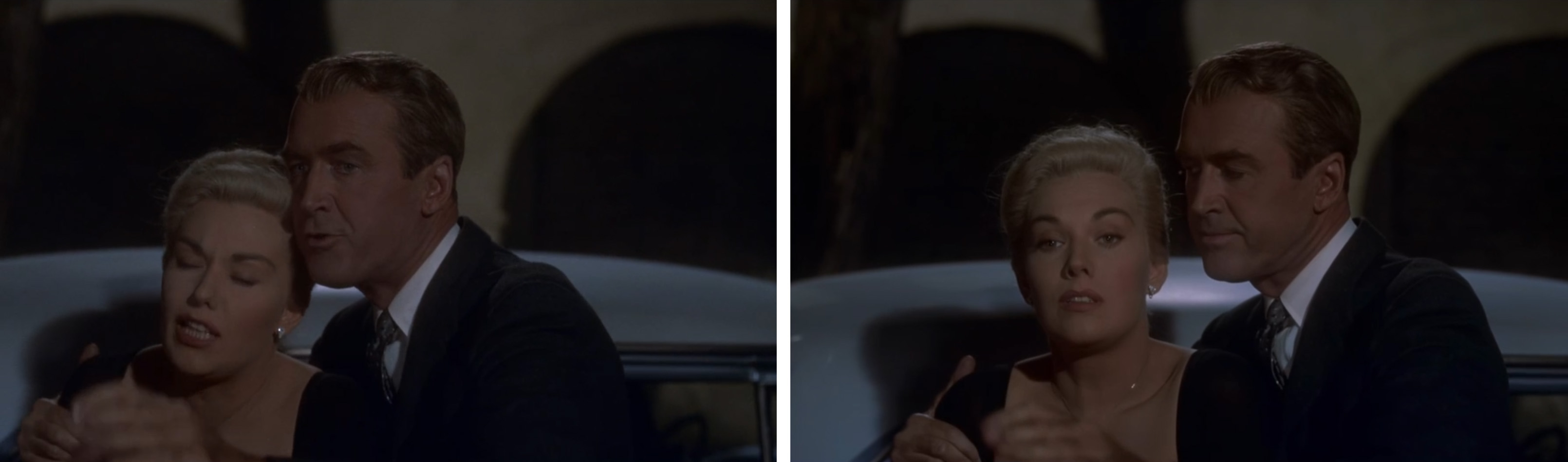}
    \caption{\textbf{Qualitative failure} in emotional rendering. In this sequence, the original film (left) captures Judy becoming visibly distraught, while the AI reconstruction (right) fails to infer this emotional shift, generating a static, unarticulated expression that divorces the character from the narrative context. Source frame: \textit{Vertigo} (1958), directed by Alfred Hitchcock. © Universal Studios.}
    \label{fig:emotive-behavior}
\end{figure}

The most overt divergences from the original cluster around specific cinematic challenges. For instance, the AI reproductions do not speak, remaining tight-lipped while the original actors deliver dialogue. At times, the generated doubles appear less expressive or project an emotion not in line with the narrative. For example, when Judy becomes visibly distraught over Scottie's arch behavior in the final act, her AI counterpart remains eerily still (Fig.~\ref{fig:emotive-behavior}). Occasionally, the model succumbs to extreme hallucinations, for example, generating unknown figures that enter and exit doors, morphing the identity of a main character beyond recognition, or producing severe deformations entirely detached from the narrative context.

\subsubsection{Where the Model Succeeds}

\begin{figure}[h!]
    \centering
    \includegraphics[width=1\linewidth]{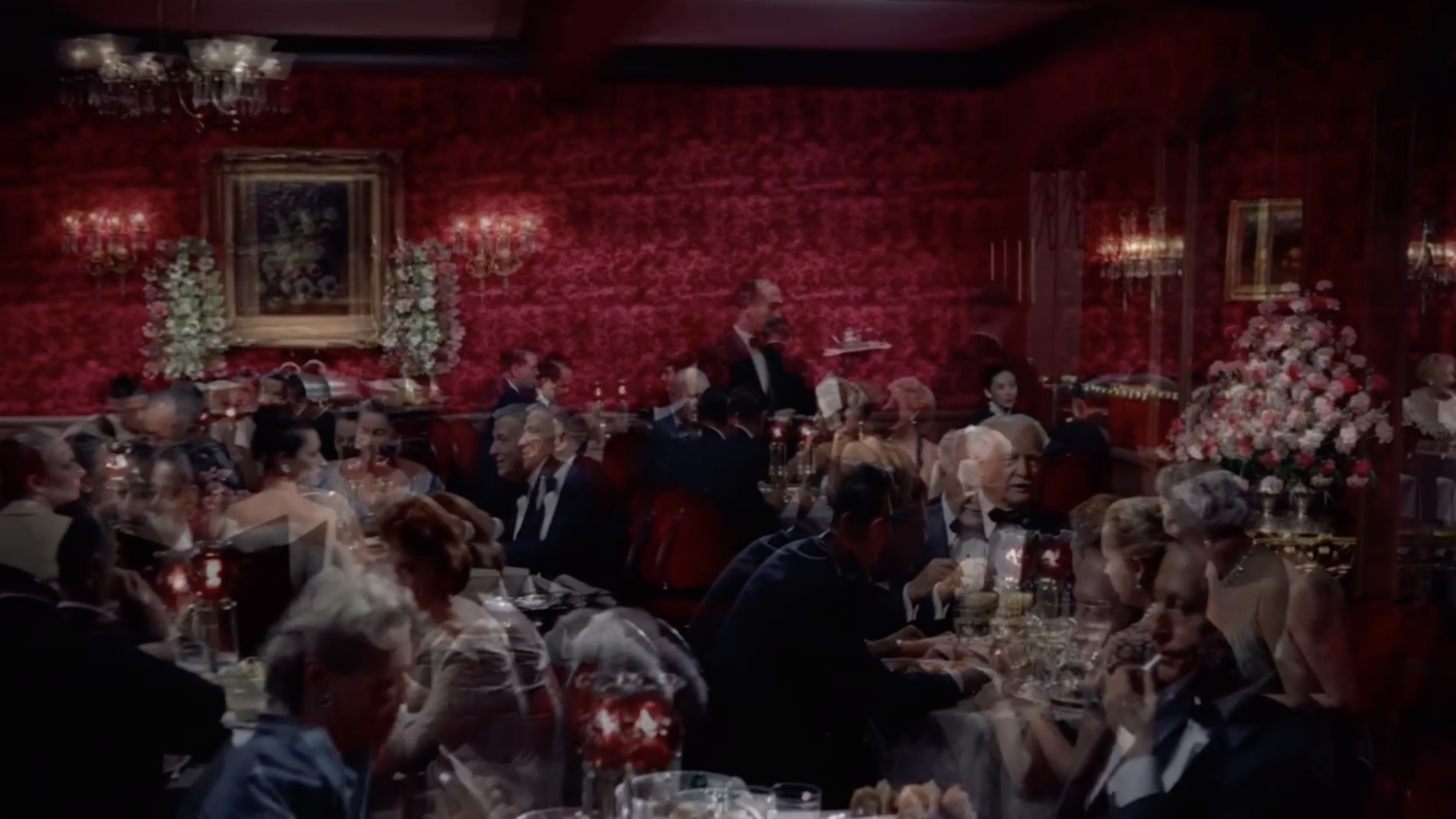}
    \caption{\textbf{Qualitative success} in high-fidelity reconstruction of a complex scene. Despite the visual density of Ernie's restaurant (featuring intricate background textures, complex geometry, and multiple moving actors), the diffusion model successfully maintains the structural integrity and spatial relationships of the scene. Derived from: \textit{Vertigo} (1958), directed by Alfred Hitchcock. © Universal Studios.}
    \label{fig:red-room-complexity}
\end{figure}

Contrasting these failures is the model's striking success in highly complex, dynamic scenes. For example, the model reconstructs the extended tracking shot through Ernie’s restaurant—a visually dense space of active diners, moving waiters and textured backgrounds—with remarkable structural fidelity, maintaining the set’s spatial geometry, the movement of actors in the scene and the camera’s focus on Madeleine (Fig.~\ref{fig:red-room-complexity}).

\subsubsection{The Plausible Middle}

\begin{figure*}[t]
    \centering
    \includegraphics[width=1\linewidth]{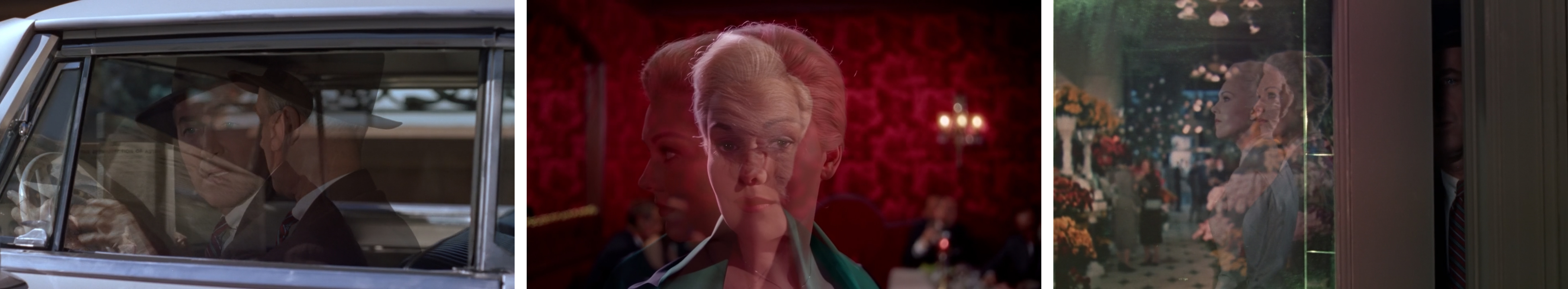}
    \caption{Representative frames from across the artwork, where the original \textit{Vertigo} and its predictive shadow are superimposed. The doubling operates on figures and spaces alike, producing a felt discorrelation at the heart of the work. Derived from: \textit{Vertigo} (1958), directed by Alfred Hitchcock. © Universal Studios.}
    \label{fig:artwork-sample}
\end{figure*}

Aesthetically, the most fascinating divergences occur not in total failure, but in these moments of near-success, when the generated double is neither a direct copy nor a complete hallucination (Fig.~\ref{fig:artwork-sample}). It exists in a plausible middle space: a reality that feels like a bad take left on the cutting room floor. When overlaid onto the original, this plausible but fictional variation has the potential to induce legitimate doubt in the viewer over what is \textit{real} and what is \textit{generated}.

\section{Discussion}

\subsection{What the Reconstruction Reveals}
\label{sec:reveals}
Across the consulted theorists, there was consistent agreement that the intervention is legible: the rhythmic oscillation between fidelity and divergence makes the nature of the AI process visible to the viewer. What is more striking is how successful the reconstruction is. Lev Manovich, surprised at the proximity of the reconstruction to the original, observed that close-ups tracked the source ``almost 100\%'' despite background figures diverging --- a fidelity he called ``a very interesting mystery'' about the model's predictive capacity. The mystery is the right framing. If interpolation distance were the governing variable, the AI's failures would scale with segment length, but that is not the case. What governs fidelity instead is what kind of image the model is being asked to produce.

Two readings of this finding are available, and they are not mutually exclusive. The first is that \textit{Vertigo} itself is well-represented in the model's training data, and the reconstruction is partly a form of lossy retrieval. The second is that classical Hollywood cinema is statistically predictable in its own right: its grammar of close-ups, two-shots, tracking movements, and continuity edits is so consistent that a model trained on the wider corpus of cinema can re-emit it from sparse anchors. Kevin Ferguson's suggestion to test the method against filmmakers with idiosyncratic visual styles, such as Godard or Cassavetes, would help disentangle these readings. What can be said now is that the visual and structural norms of classical cinema are recoverable from a small fraction of frames, and that the conventions hardest for the model to reconstruct (the Whitney spirals, the dream sequence, the late film's low-light passages) are precisely the moments that \textit{Vertigo} itself uses to mark a departure from those conventions.

These findings carry direct implications for the future of filmmaking. If a film celebrated as a pinnacle of cinema can be approximated from 2.78\% of its frames, the questions raised by the 2023 SAG-AFTRA strikes about likeness, ownership, and consent become dramatically more acute. They also reframe the long-anticipated convergence of cinema and computer animation, suggesting a near-term horizon in which entire live-action films might be produced from sparse keyframes alone, without actors, sets, or cameras.

\subsection{Inheritance and Its Limits}
While \textit{Vertigo} is a paragon of classical cinema, it is also a defining example of Mulvey's ``male gaze,'' a compelling formulation to consider in the age of machine vision \cite{mulveyVisualPleasureNarrative1975}. The model does not possess a gaze in any meaningful sense; it has no subject from which to look. What it has is a statistical disposition toward certain images, and the images it is most statistically likely to produce are those a century of cinema has rehearsed: the close-up of the blonde woman in soft focus, the tracking shot through a populated room, the conventional shot-reverse-shot of two figures in conversation. The reconstruction's pattern of successes and failures suggests inheritance: the model performs most fluently in the registers cinema has standardized, and falters in those it has not.

This continues a longer history that Shane Denson surfaced in his commentary on this work: a tradition of computational imaging protocols calibrated against idealized female bodies, most famously the Lenna test image, a 1972 \textit{Playboy} centerfold that served as the de facto benchmark for decades of digital imaging research. The point is not that the model was taught to look in a particular way. It is that certain ways of looking (for instance, certain bodies framed in certain ways) are simply more statistically likely to emerge. The historical lineage from the active male gaze of mid-century cinema to the predictive disposition of the contemporary machine model deserves closer interrogation than this paper can offer; what is clear is that there exists a continuous line between them.

This normative predisposition stands in stark contrast to the marketing rhetoric that contemporary generative systems can produce any image. They produce, with great fluency, images that conform to conventions rehearsed across more than a century of cinema. If other possibilities exist, they will require active resistance against the normative flow.

\subsection{A 21st-Century Vertigo}
The paper's claims about the present moment are expressed most effectively through the experience of watching the work itself. Viewers watch the original and its double rhythmically diverge and then harmonize at every keyframe, without necessarily being able to say, in the moment, which is which. Memory and prediction become difficult to separate. As Denson noted in his response to this work, the project provides an aesthetic correlate for the discorrelation of images from human perception more broadly, a phenomenon that, in his framing, exceeds contemporary AI and reaches back through earlier moments of compressed and predicted media. Ferguson made this lineage more explicit: MPEG video compression has long used predictive frames that interpolate between anchors; cel animation has long backgrounded what is reused; theories of cinema have long invoked saccade and suture to name the gaps the viewer's eye fills. \textit{Vertigo Vertigo} fits within this lineage, not as a paradigm shift, but as an acceleration of an existing tradition.

The artwork induces legitimate doubt over what is \textit{real} and what is \textit{generated}: an increasingly familiar panic in the digital era we may well call vertigo. We should be careful, however, of declaring a new era of unreality; we have been inundated with fictional images since the arrival of mechanically reproducible media. What is new is the rise of uniquely predictive environments, where content is algorithmically generated, personalized, and served at speeds beyond perception. The original \textit{Vertigo} itself anticipates the nested representations of the modern era: Judy performing as Madeleine performing as Carlotta, all under Scottie's obsessive voyeuristic gaze. While the desire to control this construction and approximate an ideal of authenticity is not new, it is now operationalized at scale by generative AI. This project points to an acceleration of that process: a confluence of speed, desire, and doubt that will reshape contemporary perception. Such a destabilizing discorrelation will require new sense-making artistic interventions if we are to maintain our balance.

\section{Conclusion}
This paper has shown, through an AI reconstruction of \textit{Vertigo}, that classical cinema's visual and structural norms are remarkably encoded within contemporary generative systems. Quantitatively, 73.1\% of reconstructed frames remain recognizably tethered to the original despite minimal input, with failures concentrated where the film departs from convention. Qualitatively, the work reveals a \textit{plausible middle} between copy and hallucination, a hypothetical variation that feels like a take left on the cutting room floor. Phenomenologically, the artwork produces a 21st-century vertigo: a felt discorrelation between a canonical original and its algorithmically predicted shadow.

Rather than marking a rupture, generative media extends and accelerates the historical logic of cinema, inheriting its voyeuristic gaze and desire for an artificial ideal. The problem is not whether images are real or artificial, but how these inherited structures are being reproduced, scaled, and naturalized within predictive media environments at an accelerating rate. \textit{Vertigo Vertigo} enacts this relationship between the past and present, using discorrelation as a method to make this inheritance visible.

\begin{acks}
We thank Lev Manovich, Shane Denson, Kevin L. Ferguson, Daniella Gati, and Murray Smith for their thoughtful and generous engagement with this work. Adam Cole's research is supported by the UKRI Techné Studentship, AHRC Grant reference number AH/R01275X/1.
\end{acks}

\bibliographystyle{ACM-Reference-Format}
\bibliography{PhD-bibtex}

\appendix

\section{Project Materials}
\textbf{Project Page:} \url{https://www.adamcole.studio/work/vertigo-vertigo-research-project}

\end{document}